\documentclass[twocolumn,showpacs,superscriptaddress,prl]{revtex4}

\usepackage{amssymb}

\newcommand{\cV}{\mathcal{V}}
\newcommand{\cX}{\mathcal{X}}
\newcommand{\cP}{\mathcal{P}}
\newcommand{\cN}{\mathcal{N}}
\newcommand{\cK}{\mathcal{K}}
\newcommand{\cL}{\mathcal{L}}

\newcommand{\tv}{\tilde{v}}
\newcommand{\e}{\text{echo}}
\newcommand{\beq}{\begin{equation}}
\newcommand{\eeq}{\end{equation}}
\newcommand{\bea}{\begin{eqnarray}}
\newcommand{\eea}{\end{eqnarray}}

\begin{document}

\title{Loss of quantum coherence due to non-stationary glass fluctuations}
\author{I. Martin}
\affiliation{Theoretical Division, Los Alamos National Laboratory, Los Alamos, New Mexico, 87544, USA}
\author{Y. M. Galperin}
\affiliation{Department of Physics and Center for Advanced Materials and Nanotechnology, University of Oslo, PO Box 1048 Blindern, 0316 Oslo, Norway}
\affiliation{
 Argonne National Laboratory, 9700 S. Cass Av., Argonne, IL 60439, USA}
\affiliation{
A. F. Ioffe Physico-Technical Institute of Russian Academy of Sciences,
194021 St. Petersburg, Russia}

\date{\today}

\begin{abstract}
Low-temperature dynamics of insulating glasses is dominated by a macroscopic
concentration of tunneling two-level systems (TTLS).  The distribution of the
switching/relaxation rates of TTLS is exponentially broad, which results in
non-equilibrium state of the glass at arbitrarily long time-scales.  Due to the
electric dipolar nature, the switching TTLS generate fluctuating
electromagnetic fields.  We study the effect of the non-thermal slow
fluctuators on the dephasing of a solid state qubit.  We find that at low
enough temperatures, non-stationary contribution can dominate the stationary
(thermal) one, and discuss how this effect can be minimized.
\end{abstract}


\maketitle

The main hurdle on the way to implementation of a practical quantum computer
lies in the design of a quantum system that is well isolated from the
detrimental influences of environment, but at the same time accessible enough
to allow coherent manipulation necessary to perform quantum computation.  These
are typically conflicting requirements since any external device that performs
manipulation on a qubit is also a likely source of noise, which leads to
decoherence. For example, metallic gates, which are commonly used to control
solid-state qubits such as a superconducting Cooper pair box  \cite{nakamura}
or a donor spin in a semiconductor \cite{kane}, as well as in ion traps
\cite{itrap}, are a source of electro-magnetic noise.  This noise is generated
by the low-energy particle-hole excitations in the metal, which in some cases
is manifested as the Johnson-Nyquist Ohmic noise \cite{ingold_nazarov}.

Decoherence can also be induced by the low energy degrees of freedom in the
insulator surrounding the qubit.  Recently, a number of studies were performed
to analyze the role of the two-state fluctuators located in the insulators on
the qubit dephasing and relaxation
\cite{paladino,galperin,faoro,grishin,shnirman,sousa}. Such fluctuators
naturally lead to charge noise. It is usually assumed that the fluctuators are
in a thermal equilibrium, which is maintained via their interaction with the
external environment and among themselves. However, very often the insulators
used in the solid state qubits -- Si$_3$N$_4$, SiO$_2$ or Al$_2$O$_3$ -- are
amorphous. It is well known that in amorphous materials, a large number of
two-level systems exist with arbitrarily long switching times. In particular,
they lead to the logarithmic time dependence of the specific heat
\cite{zeller}. Some of these slow two-level systems are frozen in high-energy
states with energies significantly exceeding the nominal thermal energy and
when they eventually relax to the ground state they stay there indefinitely. In
this Letter we study the effects of such over-cooled two-level systems on the
qubit dephasing. We find that for a typical glass they can give a significant
contribution to the total dephasing rate, up to 1 GHz right after the system
cool-down.

The most common type of two level-systems (TLS) in structural glasses
are the \textit{tunneling} TLS, or TTLS.  Their switching rate
$\gamma$ is determined by  either over-the-barrier thermally-activated
tunneling, or under-the-barrier quantum tunneling. Thus the tunneling
rate is an exponential of a uniformly distributed parameter
characterizing the barrier strength. This generically results in the
probability distribution $\cP(\gamma) =
[\gamma\log(\gamma_{\max}/\gamma_{\min})]^{-1}$.  An ensemble of TTLS
with the energy splittings less than the thermal energy, $E_i \lesssim
k_B T$, naturally leads to the 1/f noise \cite{dutta}, typical of
glasses.  The main contribution of this noise is concentrated at low
frequencies $\hbar \omega < k_B T$; therefore, it primarily affects
the qubit dephasing (not relaxation), and moreover can be treated as a
classical noise \cite{galperin,shnirman}.  The number of thermally
active two-level systems scales linearly with the temperature
(assuming flat distribution $\cP(E)$ at small level splittings $E_i$),
which for very low temperature would seem to imply that dephasing
should disappear.  Here we argue, however, that at low enough
temperature, slow \textit{non-thermal} two-level fluctuators will
dominate over the contribution  from the thermal ones. That is because
the number of non-thermal fluctuators is proportional not to the
temperature but to the relevant band width $T_0$ of TTLS, which can be
up to 20~Kelvin \cite{soft}.

The electromagnetic noise generated by the glassy metastable
fluctuators can affect a variety of qubit architectures.  However,
for concreteness, here we consider the effect of the electric field
fluctuations generated by the TTLS switching on a superconducting
qubits.  Electric field fluctuations couple to the qubit charge.  In
the qubit charge basis, the Hamiltonian in the presence of the charge
noise $\cX(t)$ is
\beq H = -\frac{E_C}{2} \sigma_z -\frac{E_J}{2}
\sigma_x + \frac{\cX(t)}{2}\sigma_z \,, \eeq
where $\sigma_z$ and
$\sigma_x$ are the Pauli matrices acting on the space of the qubit
states $\{0,1\}$ and $E_C$ and $E_J$ are the qubit charging and
Josephson energies.  Diagonalizing the noise-free part of the
Hamiltonian and keeping only the \textit{longitudinal} coupling to the noise
(assuming that its contribution is negligible at the qubit frequency
$\hbar\omega = E\equiv\sqrt{E_C^2 + E_J^2}$ and thus cannot
effectively lead to qubit relaxation) the Hamiltonian becomes
\beq \label{eq:H2}
H =
-\frac{E}{2}\, \sigma_z +\frac{E_C}{2E}\, \cX(t)\sigma_z.  \eeq
Note that TTLS can be also present in the insulating parts of
Josephson contacts. Their switching then leads to fluctuations of the
Josephson energy $E_J$ \cite{Kozub,Harlingen}, the relevant
contribution to the Hamiltonian~(\ref{eq:H2}) being
$(E_J/2E)\cX(t)\sigma_z$. This contribution is important only near
degeneracy point, $E_C=0$, where it can be taken into account in a
similar way.

For slow fluctuators the noise can be treated classically and it is a sum of
individual contributions $\cX(t) = \sum{v_i z_i(t)}$, where $v_i$ is the
strength of the coupling between the qubit and the $i^{\rm th}$ fluctuator, and
$z_i(t)$ describes the time evolution of the fluctuator.  To single out the
effect of non-thermal fluctuators, we neglect here the thermally excited
fluctuators, such that $E \lesssim k_B T$.  Without loss of generality, we set
for the excited fluctuator $z = 1$ and for the ground state $z =0$.

The dephasing effect of a particular realization of the fluctuators' dynamics
on the qubit can be defined as
\bea\label{eq:fid}
 e^{i\frac{E_C}{E}\int_0^t{dt' \cX(t')}}=\prod_i
e^{i\int_0^t{dt' \tv_i z_i(t')}}. \eea It describes the phase drift between the
0 and 1 sates of the qubit due to the random switching events in the
environment, as would be measured in a {\em free induction decay} (FID)
experiment~\cite{slichter}. Here $\tv \equiv ({E_C}/{E})v$.  The dependence on
the biasing point ${E_C}/{E}$ is routinely used to disentangle the charge noise
acting on qubit from the phase/flux noise. To determine the statistical
dephasing, we need to average over the time evolutions $z_i(t)$, the parameters
of the fluctuators (the switching rates $\gamma$), and the coupling strengths
$v$. Assuming uncorrelated fluctuators, we can perform the evolution averaging
independently for all $z_i$.  For an exponential relaxation from the excited
state with switching rate $\gamma_i$, $\cP(1|1) = e^{-\gamma_i |t|}$, the
$average$ single-fluctuator contribution to dephasing is \bea\label{eq:psi_fid}
\psi_i^{\rm FID}(t) &=& \int_0^t{dt' \gamma_i
e^{-\gamma_i t'} e^{i\tv_i t'} } + e^{-\gamma_i t}\,e^{i\tv_i t} \nonumber\\
 &=& \frac{i\tv_i\, e^{i\tv_i t- \gamma_i t} - \gamma_i}{i\tv_i-\gamma_i}\, .
\eea This expression corresponds to the average dephasing factor
due to the $i^{\rm th }$ fluctuator, which one would obtain in an experiment
repeated with the identical initial conditions (fluctuator $i$ in the excited
state at $t=0$). Accordingly, the dephasing factor due to an ensemble of
uncorrelated fluctuators is
\bea\label{eq:Psi} \Psi(t) = \prod_i \psi_i(t) \, . \eea

When many fluctuators interact with the qubit simultaneously, one can evaluate
the dephasing factor by performing the average over the distribution of
$\{\tv_i, \gamma_i \}$.  We make the following assumptions: (i) Fluctuators are
distributed in 3D according to the Poisson distribution, $dN(r) = n 4\pi r^2
dr$, where $n$ is the concentration; (ii) The coupling to qubit is of dipolar
form $v(r) = A/r^3$, where A can be either positive or negative; (iii) There is
no correlation between $\gamma$ and $v$; (iv) the distribution function for the
switching rate has the tunneling form $P(\gamma) = ({\cal L} \gamma)^{-1}$,
where $\cal L = \ln(\gamma_{\max}/\gamma_{\min})$.   Under these assumptions,
the full distribution function (normalized to the total number of fluctuators
${\cal N} = n {\cal V}$),
\begin{equation}\label{eq:P}
{\cal P}(v,\gamma) ={\cal N}\, \frac{\eta}{\cal L}\, \frac 1 {v^2\gamma}
e^{-\eta/|v|}, \quad \eta = \frac{4 \pi A}{3 {\cal V}} \, .
\end{equation}
The physical meaning of $\eta$ is the coupling between the qubit and the
fluctuator located at the boundary of the volume ${\cal V}$ (the ``weakest"
fluctuator).

To evaluate dephasing factor in Eq.~(\ref{eq:Psi}), we use the
Holtsmark procedure~\cite{chandrasekhar} for averaging the
logarithm of the product,
\begin{equation}\label{eq:k_fid}
{\cal K}^{\rm FID}(t) = -\ln \Psi^{\rm FID}(t) \approx \int dv d\gamma \, {\cal
P}(v,\gamma)g^{\rm FID}(v,\gamma|t)\, ,
\end{equation}
where $g^{\rm FID}(v,\gamma|t) \equiv 1 - [{\psi^{\rm FID}_{+v}(t)
+ \psi^{\rm FID}_{-v}(t)}]/{2}$ includes averaging over the sign
of coupling $v$. According to our assumptions, the distribution
function does not depend on the sign of $v$. Thus we can perform
the sign average in $\psi$ and limit the integration over $v$ to
positive values only. From Eqs. (\ref{eq:psi_fid}) and
(\ref{eq:P}), we obtain for $t >0$
\begin{equation} \label{eq:K_fid}
\cK^{\rm FID}(t)=\tv_{\rm typ
}\int_{\gamma_{\min}}^{\gamma_{\max}}\frac{d\gamma}{\gamma^2}\,
  \left( {1-e^{-\gamma t}} \right)\, .
  \end{equation}
For $\gamma_{\max}t \ll 1$ this yields,
\begin{equation} \label{eq:k_fid01}
 \cK^{\rm FID}(t \ll \gamma_{\max}^{-1})= \cL \,\tv_{\text{typ}}t\, ,
\end{equation}
while in the other regime
\begin{equation}
  \label{eq:k_fid02}
 \cK^{\rm FID}(\gamma_{\max}^{-1}\ll t \ll  \gamma_{\min}^{-1})=
 \tv_{\text{typ}}t\, \ln \left( \frac{1}{\gamma_{\min}t}\right) \, .
\end{equation}
The characteristic coupling strength that determines the decay of coherence is
defined here as \mbox{$\tv_{\rm typ} = ({E_C}/{E})2\pi^2A \cN/3\cV \cL$}.
Note that both
short and long-time decay laws (\ref{eq:k_fid01}) and
(\ref{eq:k_fid02}) are nearly exponential.

The free induction decay is the simplest measure of dephasing.  However, even a
fluctuator that never switches on the timescale of the experiment will give a
contribution to FID dephasing if we use definition (\ref{eq:fid}). Moreover,
experimentally it is impractical to reset (``re-initialize") the states of the
fluctuators.  This dependence on the fluctuator initial-state can be easily
eliminated by measuring the dephasing $relative$ to the initial qubit level
splitting, which includes contributions from the two-level fluctuators. The
result is the following definition for dephasing \cite{Laikhtman}
\bea\label{eq:fid2}
 e^{i\frac{E_C}{E}\int_0^t{dt' [\cX(t')-\cX(0)]}}
\eea
which yields
\begin{equation} \label{eq:k_rel}
\cK^{\text{rel}}(t)=-\tv_{\rm
typ}\int_{\gamma_{\min}}^{\gamma_{\max}}\frac{d\gamma}{\gamma^2}\,
  \left( 1-\gamma t-{e^{-\gamma t}} \right)\, ,
\end{equation}
and in the short and long-time limits
\begin{equation}
\label{eq:k_rel01}
 \cK^{\text{rel}}(t)= \left\{ \begin{array}{ll}
     \tv_{\rm typ}\gamma_{\max}t^2\,, & t \ll \gamma_{\max}^{-1}  \\
    {\ln ( \gamma_{\max}t)}\,\tv_{\rm typ}
 t\,, & \gamma_{\max}^{-1}\ll t \ll  \gamma_{\min}^{-1} \,.
 \end{array} \right.
\end{equation}
With this definition, the long-time dephasing is similar to FID, however, in
the short time limit the dephasing is quadratic in time and thus dramatically
suppressed.

Example with the relative dephasing shows that subtraction of the systematic
qubit frequency shift dramatically reduces dephasing.  However, direct
implementation of such subtraction protocol is not possible, except for the
slowest fluctuators that remain frozen from one experimental run to another.
Alternatively, a similar subtraction effect can be obtained by means of {\em
spin echo} \cite{slichter} protocol.  In the echo experiment, the static phase
accumulation is eliminated by introducing $\pi$ pulse on qubit at time $t/2$
and performing measurement at time $t$,
\bea\label{eq:echo}
 e^{i\frac{E_C}{E}\int_0^{t/2}{dt' \cX(t')} - \int_{t/2}^{t}{dt' \cX(t')}}.
\eea
Integration over $v$ distribution now yields
\begin{equation}
  \label{eq:k_echo}
  \cK^{\e} (t)=\tv_{\rm typ}\int_{\gamma_{\min}}^{\gamma_{\max}}\frac{d\gamma}{\gamma^2}\,
\left(1+e^{-\gamma t}-2 e^{-\gamma T/2}  \right)
\end{equation}
and
\begin{equation}
\label{eq:k_echo01}
 \cK^{\e}(t)= \left\{ \begin{array}{ll}
   \frac{1}{4} \tv_{\text{typ}}\gamma_{\max} t^2\,, & t \ll \gamma_{\max}^{-1}  \\
    \tv_{\text{typ}} t\,\ln 2, &\gamma_{\max}^{-1}\ll t \ll  \gamma_{\min}^{-1}
    \, .
 \end{array} \right.
\end{equation}
Indeed, the result is qualitatively similar to the one for the
protocol (\ref{eq:fid2}).  Notice also that the long-time
dephasing for all protocols, including FID, is the same.

In thermal equilibrium, the maximum switching rate $\gamma_{\max}(T)$ is
determined by the interaction between a TTLS with electrons or phonons.  The
estimates have been obtained for amorphous metals~\cite{black} and dielectric
glasses~\cite{jackle}. At a given temperature $T$, these estimates can be are
$\gamma_{\max}(T) \approx \chi T/\hbar$ where $\chi =0.01 - 0.3$ for amorphous
metals, and $\gamma_{\max}(T) \approx T^3/\hbar T_c^2$ where $T_c =15-30$ K
(depending on the elastic parameters) for dielectric glasses.  The minimal
switching rate, $\gamma_{\min}$, is actually not known. Logarithmic heat
release from structural glasses was observed during many hours, see e.~g.
Ref.~\onlinecite{Parshin-Sahling} and references therein. The low-frequency
noise in disordered materials has $1/f$ spectrum down to any observable
frequencies.

The non-stationary case considered here, however, requires careful
interpretation of $\gamma_{\max}$ and $\gamma_{\min}$.  The upper limit of the
switching rate, $\gamma_{\max}$, is determined by the time delay, $\tau_d$,
since the preparation of the system (e.~g., since the cool-down, or some other
strong manipulation that can reset the fluctuators).  Indeed, the fluctuators
with $\gamma \gtrsim 1/\tau_d$ are likely to decay before the the start of the
measurement. This naturally leads to a cut-off $\gamma_{\max} = 1/\tau_d$.  The
simplest assumption about the lower cut-off on relaxation rate is that it is
the same as in the equilibrium case.  However, it can also can also depend on
history. For instance, if the system was warmed-up to temperature $T_0$ for a
period of time $\tau_p$, then the fluctuators with energies less than $T_0$,
and relaxation rates faster than $1/\tau_p$ will get ``recharged."  To include
this possibility, we define $\gamma_{\min} = 1/\tau_p$.

Since the measurement time in a typical experiment is less than $\tau_d$ we can
concentrate only on the ``short-time" limit, $t \ll \gamma_{\max}^{-1}$. Then,
\begin{equation}
\label{eq:k_shortt}
 \cK(t \ll\tau)= \left\{ \begin{array}{ll}
     \tv_{\rm typ} t\,, &\quad{\rm FID}  \\
   \frac{\tv_{\rm typ}}{4\tau_d} t^2\,, &\quad{\e}\, .
 \end{array} \right.
\end{equation}
It was shown recently that the short-time limit of the echo decay has the
desirable {\em self-averaging} property, that is the ensemble average
(calculated here) indeed corresponds to the typical result from the a single
sample, averaged over repeated experimental runs \cite{galperin,schriefl}. On
the other hand, FID and the long-time limit of echo do not self-average due to
strong mesoscopic fluctuations in the positions of the nearest
fluctuator(s)\cite{schriefl}.

We now crudely estimate the magnitude of dephasing assuming that the qubit
environment is an insulating glass, such as SiO$_2$ or Al$_2$O$_3$.  The
concentration of TTLS's in such systems is about $\cP_0 \equiv \cN/\cV \cL \sim
10^{32}-10^{33}$ cm$^{-3}$erg$^{-1}$~\cite{soft}. The energy distribution of
TTLS is uniform, with the band width of about 20 Kelvin.  The band width is
much larger than the typical temperature $10-50$ mK in the qubit
dephasing/relaxation experiments, and therefore only a small fraction of TTLS
are thermally excited and contribute to the telegraph noise capable of
dephasing the qubit.  On the other hand, the fraction of metastable fluctuators
that do not decay during the time $\tau_d$ since the cool-down is $\cN = \cP_0
\cV T_0\ln(\tau_p/\tau_d)$. Here $T_0$ is  the relevant bandwidth of the
nonequilibrium fluctuators, e. g., the temperature before final cool-down.

The coefficient in the dipolar interaction for a TTLS that interacts with its
mirror image  on the surface of the qubit is $A\sim e^2 d^2$, where the typical
dipole moment, $ed$,  corresponds to one electron charge displaced by $d \sim
1$ {\AA}. Assuming the bandwidth of the frozen TTLS to be $T_0 \sim 1$ K we
obtain as a crude estimate, $v_{\text{typ}}/\hbar \sim 2\pi^2 e^2d^2 \cP_0 T_0
\sim 10^{8}-10^{9}$ s$^{-1}$. Introducing now the dephasing time as the time
during which the log of the average phase factor becomes equal to one,
$\cK(\tilde{T}_2) = 1$, we find $\tilde T_2^{\rm FID} \sim
(E/E_C)v_{\text{typ}}^{-1}\sim (E/E_C)\cdot (10^{-8}-10^{-9})$ s and $\tilde
T_2^{\rm echo} \sim \sqrt{\tau_d\, \tilde{T}_2^{\text{FID}}}$.

It is instructive to compare these results with the contribution to dephasing
from the thermally excited fluctuators. Let us make this comparison for the
echo signal. The effective number of thermal fluctuators is less than number of
the frozen ones by the factor $\sim T/T_0 \ll 1$.  Since $v_{\rm typ}$ is
proportional to concentration,
$v_{\text{typ}}^{\text{frozen}}/v_{\text{typ}}^{\text{thermal}}
 \sim T_0/T\gg 1$.  For thermal fluctuators~\cite{galperin}
 $$\frac{1}{\tilde{T}_2^{\e}} =\min \left\{v_{\text{typ}}^{\text{thermal}},
  \sqrt{v_{\text{typ}}^{\text{thermal}}\cdot\gamma_{\max}(T)}\right\}
  \, .
 $$
At experimentally relevant temperatures, in dielectric glasses one can expect
$\gamma_{\max}(T)\ll  v_{\text{typ}}^{\text{thermal}}$.  For example, at $T =
50$~mK, $v_{\text{typ}}^{\text{thermal}}\sim 10^6$~s$^{-1}$ and
$\gamma_{\max}(T)\sim 10^3$~s$^{-1}$.  Therefore,
  $$
  \frac{\tilde{T}_2^{\text{frozen}}}{\tilde{T}_2^{\text{thermal}}}
  \sim \left(\frac{T}{T_0}\right)^{1/2}\! \sqrt{\gamma_{\max}(T)
  \tau_d}\, .
  $$
  Assuming that $\gamma_{\max}(T)=T^3/T_c^2 \hbar$ we get
 \begin{equation} \label{eq:0001}
 \frac{\tilde{T}_2^{\text{frozen}}}{\tilde{T}_2^{\text{thermal}}}
 \sim \frac{T^2}{T_c}\, \left(\frac{
 \tau_d}{\hbar T_0}\right)^{1/2}
 \, .
    \end{equation}
This relation shows that at
$$\tau_d \ll  \hbar T_0T_c^2/T^4 $$
frozen fluctuators provide more decoherence than thermal ones.  Assuming $T_0
=1$ K, $T_c=20$ K, $T=50$ mK, we conclude that the crossover occurs at
$\tau_d\sim 10^{-2}$~s.  While this delay time is much shorter that a typical
time from a cool-down to an experiment, it is much longer than the typical
experimental inverse repetition rate.\cite{nakamura}  In case the manipulation
procedure includes application of significant bias or gate voltages, e.g. at
the measurement stage, this can lead to ``charging-up'' of the non-thermal
fluctuators.  In particular, a temporary swapping of ground and excited states
of a fluctuator can be induced by the gate electric fields.

In conclusion, we analyzed the effect of slow non-thermal metastable glass
fluctuators on qubit dephasing.  We found that at low enough temperatures, the
non-thermal dephasing can exceed the contribution of the thermal fluctuators.
This effect can be reduced by appropriately designing manipulation sequences in
order to avoid re-exciting of the metastable fluctuators and by choosing
sufficiently long delay time between consequent manipulations.

\acknowledgments We would like to thank V. Kozub and A. Shnirman for useful
discussions.  This work was supported by the U. S. Department of Energy Office
of Science through contracts No. W-7405-ENG-36 and No. W-31-109-ENG-38.

\end{document}